The challenge of engaging *all* students via self-paced interactive e-learning tutorials for introductory physics

Seth DeVore, Emily Marshman and Chandralekha Singh

Department of Physics and Astronomy, University of Pittsburgh, Pittsburgh, PA 15260, USA

**Abstract:** As research-based self-paced e-learning tools become increasingly available, a critical issue educators encounter is implementing strategies to ensure that all students engage with them as intended. Here, we discuss the effectiveness of research-based e-learning tutorials as self-paced learning tools in large enrollment brick and mortar introductory physics courses. These interactive tutorials were developed via research in physics education and were found to be effective for a diverse group of introductory physics students in one-on-one implementation. Instructors encouraged the use of these self-paced tools in a self-paced learning environment by telling students that they would be helpful for solving the assigned homework problems and that the underlying physics principles in the tutorial problems would be similar to those in the in-class quizzes (which we call paired problems). We find that many students, who struggled in the courses in which these adaptive e-learning tutorials were assigned as a self-study tool, performed poorly on the paired problems. In contrast, a majority of student volunteers in one-on-one implementation greatly benefited from the tutorials and performed well on the paired problems. The significantly lower overall performance on paired problems administered as an in-class quiz compared to the performance of student volunteers who used the research-based tutorials in one-on-one implementation suggests that many students enrolled in introductory physics courses did not effectively engage with the self-paced tutorials outside of class and may have only used them superficially. The findings suggest that many students in need of out-of-class remediation via self-paced learning tools may have difficulty motivating themselves and may lack the self-regulation and time-management skills to engage effectively with tools specially designed to help them learn at their own pace. We conclude by proposing a theoretical framework to help students with diverse prior preparations engage effectively with self-study tools.

## 1. INTRODUCTION

Effective use of e-learning tools for self-paced learning can provide a variety of students an opportunity to learn using an approach that allows each student to make progress at a pace that is commensurate with their prior knowledge [1-5]. Many instructors provide web-based tools to their students to encourage "self-study" outside of class, even in brick and mortar classes. Since there is limited time available in the classroom to address the needs of students with different prior preparations, research-based self-study tools can provide a valuable opportunity to supplement in-class learning for all students. They have the potential to help students learn to think in an expert-like manner while engaging in problem solving and can expose students to concepts in a way that scaffolds learning [6-10].

Some tools available to students are adaptive in that they adapt to students with different prior knowledge and skills. For example, an adaptive learning tools may provide more scaffolding support to a student who is struggling than others in the same course. Moreover, adaptive e-learning tools that are designed via research can be particularly beneficial because they can help a variety of students with different prior preparations and allow them to learn at their own pace [11-15].

However, an issue that instructors often encounter is achieving appropriate student engagement with these self-paced learning tools especially among those who are struggling with the course material and are in need of remediation. In particular, many students may have difficulty motivating themselves and may lack self-regulation and time-management skills which are critical for effectively engaging with self-study tools [16-19]. Without sufficient help for developing these skills and incentives to motivate them, students may not follow the guidelines for effective use of the self-paced adaptive e-learning tools. The ineffective use of the tool can significantly reduce its effectiveness. It is therefore important to investigate how students engage with self-study tools, as well as how to best incentivize their use.

In this research, we focus on the effectiveness of encouraging introductory physics students in large brick and mortar classes to use adaptive, interactive e-learning tutorials as a self-study tool to help them with their homework and quizzes. These tutorials were developed using research in physics education and were refined through an iterative process including feedback from students and instructors [12-15]. They are designed to aid students with diverse backgrounds via a guided approach to learning, in which guiding questions provide scaffolding support to help students learn physics concepts and develop useful skills. The tutorials are adaptive in that students get feedback when they select an incorrect answer, which is based upon their conceptual difficulty. Student learning is evaluated by their performance on "paired problems", which were administered to students as a weekly recitation quiz and emphasize concepts covered in the associated tutorial. The goals of the study are as follows:
- Determine the effectiveness of adaptive, interactive e-learning tutorials for a diverse group of introductory physics students at a large research university in one-on-one implementation in which researchers ensured that the tutorials were used as intended
- Determine the effectiveness of the e-learning tutorials for a diverse group of introductory physics students in brick-and-mortar introductory physics courses in which researchers had no control over how the tutorials were used by the students
- Compare the performance of the students who worked on the e-learning tutorials in a one-on-one interview vs. those who used them as a self-study tool
- Develop a theoretical framework to interpret possible differences in the performance of the students who worked on the e-learning tutorials in a one-on-one interview vs. those who used them as a self-study tool that can be useful for effective implementation of self-paced learning tools in the future.

Below, we first describe the structure of the interactive adaptive e-learning tutorials. Then, we describe the investigation of the effectiveness of these tutorials as a self-study tool in one-on-one implementation. Next, the investigation focusing on the implementation of the tutorial as a self-study tool as a part of traditionally taught large introductory physics classes is discussed. We then compare the effectiveness of these self-paced learning tools in large classes vs. their effectiveness in one-on-one implementation. Then, we discuss how inadequately incentivized self-paced learning tools may not have a positive impact on learning even if they are developed via research. Finally, we propose a theoretical framework that includes considerations critical for helping students with diverse prior backgrounds and preparations benefit from adaptive, interactive self-study tools.

## 2. TUTORIAL DEVELOPMENT AND STRUCTURE

The development of the e-learning tutorials is guided by a cognitive apprenticeship learning paradigm [20, 21] which involves three essential components: modeling, coaching, and weaning. In this approach, "modeling" implies that the instructor demonstrates and exemplifies the skills that students should learn (e.g., how to solve physics problems systematically). "Coaching" means providing students opportunity, guidance and practice so that they are actively engaged in learning the skills necessary for good performance. "Weaning" consists of reducing the support and feedback gradually so as to help students develop self-reliance.

The e-learning tutorials are developed to model and coach students to learn physics content and develop useful skills and wean students as they develop self-reliance. Each e-learning tutorial starts with an overarching problem which is quantitative in nature. Figure 1 is an example of one of these overarching problems. Before working through a tutorial, students are asked to attempt the problem to the best of their ability. The tutorial then divides this overarching problem into a series of sub-problems, which take the form of research-guided conceptual multiple-choice questions. These sub-problems help students learn effective steps for successfully solving a physics problem, e.g., analyzing the problem conceptually, planning the solution and decision making, implementing the plan, and assessing and reflecting on the problem solving process. The alternative choices in these multiple-choice questions elicit common difficulties students have with relevant concepts. Incorrect responses direct students to appropriate help sessions in which students are provided suitable feedback and explanations both conceptually and with diagrams and/or appropriate equations to learn relevant physics concepts. Correct responses to the multiple-choice questions advance students to a brief statement affirming their selection followed by the next sub-problem. In addition to the tutorial problem statement in Figure 1, the investigation described here was conducted on two other tutorials. In the Newton's second law tutorial, students are provided a set of three blocks on an inclined plane connected via strings to each other and being pulled up the incline. They are asked to determine the acceleration of the middle block and the tensions in all strings. In the conservation of mechanical energy/work-energy theorem tutorial students are provided a problem in which they must use both conservation of mechanical energy and work-energy theorem for two sub-problems of the problem. In this problem, they are asked to determine the safety of a stunt in which a man is shot out of a spring-loaded cannon and onto an airbag. For the first sub-problem involving mechanical energy conservation, students are provided the initial compression of the spring, and various heights in order to be able to figure out the changes in the gravitational potential energy so that they can find the speed of the person right before he falls on the airbag. For the second sub-problem involving the work-energy theorem, they are provided the thickness of the airbag and the average force the airbag exerts on the person in order to figure out whether the person stops before the airbag is fully compressed (if that is the case, the person is safe).

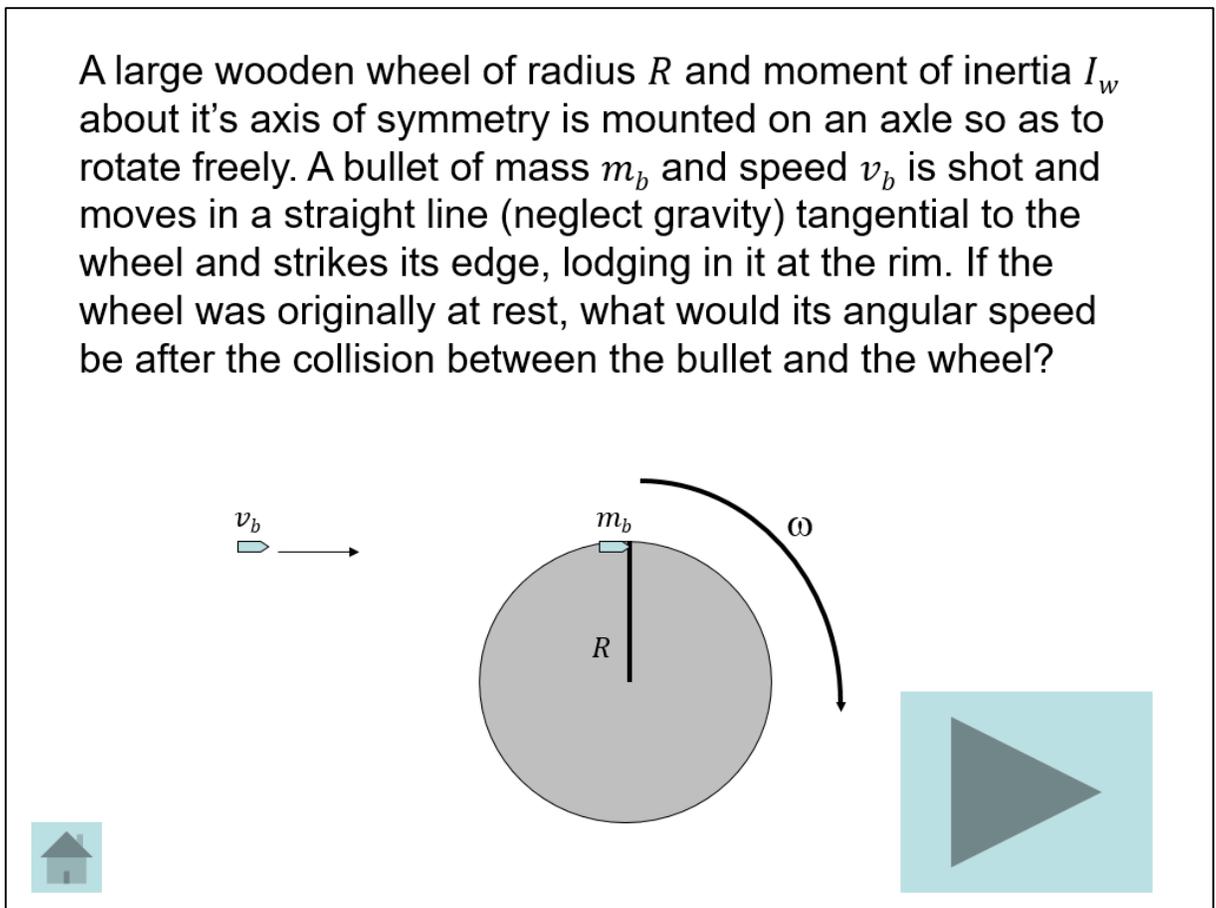

Figure 1: The overarching problem in the conservation of angular momentum tutorial.

Figure 2 shows examples of sub-problems in two of the e-learning tutorials. The top two image in Figure 2 are from the Newton's second law tutorial and provides students an opportunity to determine which free body diagram is correct for a system of three blocks that are in contact resting on an inclined plane with a force applied upwards along the plane. The tutorials are adaptive in that they provide feedback and help to students if they select an incorrect answer to a sub-problem which is commensurate with their difficulty. For example, in Fig. 2, if students select option A, they are provided with help that focuses on the fact that the plane applies a normal force. If they select option C (as shown in Figure 2), the help focuses on the fact that the normal force should point perpendicular to the surface, and, similarly, if they select option D, the help focuses on the fact that the gravitational force points vertically down, instead of pointing in the direction perpendicular to the surface. The bottom two images in Figure 2 are from the conservation of energy tutorial. Students are asked to determine which forms of energy the spring-Dave-earth system possesses before the spring is released. If students select option A, they learn that the system possesses no kinetic energy because Dave has zero initial speed. If they select option B, they learn that the system does possess spring potential energy but that Dave also started at some initial height above the reference height so there is some gravitational potential energy. Similarly, if students select option C, they learn that the system possesses gravitational potential energy but the spring was initially compressed so there is non-zero elastic potential energy also. The feedback students obtain when they select the correct answer (option D) is shown in Figure 2 and confirms that the system has both elastic and gravitational potential energy at the moment in question.

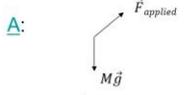

Figure 2: Examples of two sub-problems (left) and two associated responses when students select a choice (right) from tutorials focusing on Newton's second law (top) and conservation of energy/work-energy theorem (bottom). For the sub-problems, students can either click on a particular option in the multiple-choice or click on the home button in order to access any of the previous sub-problems and associated help.

After students work on the sub-problems, they answer several reflection sub-problems. These reflection sub-problems focus on helping students reflect upon what they have learned and apply the concepts learned to different contexts. If students have difficulty answering the reflective sub-problems, the tutorial provides further assistance and feedback in the form of a review of the effective problem solving approach. Thus, this interactive, adaptive e-learning tool does not merely model or exemplify a systematic approach to problem solving, it engages students actively in the use of this systematic approach and provides feedback and guidance based on their need. The tutorial will provide less scaffolding if students become more confident in solving the later sub-problems on their own without help.

Each e-learning tutorial problem is matched with paired problems that use similar physics concepts but which are somewhat different in context. Students can be given these paired problems as quizzes so that they learn to de-contextualize the problem solving approach and concepts learned via the e-learning tutorial. The paired problems also play an important role in the weaning part of the learning model and ensure that students develop complete self-reliance and are able to solve problems based upon the same concepts without any guidance. Students' performance on the paired problems after they work on the tutorial was used to evaluate the effectiveness of each tutorial. One of the paired problems posed in association with the tutorials is as follows:

> *A 20kg boy stands on a small stationary (at rest) merry-go-round near the edge of the merry-go-round. The total moment of inertia of the system of merry-go-round with the boy on it about the center is 120kg m².  The boy at the edge of the merry-go-round (radius of 2m) jumps off the merry-go-round in a tangential direction with a liner speed of 1.5m/s. What is the angular speed of the merry-go-round after the boy leaves it?*

The other paired problem associated with this tutorial and the ones related to the other two tutorials used in this investigation are available for download online [22, 23].

Twenty such e-learning tutorials were developed, which cover many topics in introductory physics related to mechanics, electricity, and magnetism. In this investigation, we focus on the effectiveness of three of these tutorials on introductory mechanics. The three adaptive e-learning tutorials selected for this research study were developed to improve student understanding of physics principles which are central in an introductory mechanics course: application of Newton's second law, conservation of energy/work-energy theorem, and conservation of angular momentum. All three adaptive e-learning tutorials were developed using the protocol discussed above. First, a quantitative problem that requires use of these physics principles was selected. Each tutorial problem was chosen to be somewhat more difficult than a typical introductory level physics homework problem on the same physics principle (these problems were used for quiz in introductory physics courses at the same university so their difficulty level was known). This level of difficulty was chosen so that the problems could not be solved using a plug and chug approach and would have enough depth to be able to help students learn an expert-like problem solving approach. Then, a cognitive task analysis was performed by three graduate student researchers and one professor (all physics education researchers) to break down each tutorial problem into a series of sub-problems dealing with different stages of problem solving that must be answered to solve the tutorial problem. Each sub-problem was then posed as a multiple choice question. The incorrect options for each multiple choice question were chosen to emphasize common difficulties uncovered by having introductory physics students solve similar problems in an open-ended format. Explanations for each multiple choice option were written and refined, based on one-on-one student interviews, to reinforce student understanding of the reasoning behind the options given and to aid them in repairing their knowledge structure when they select an incorrect option. Using this approach, the initial drafts of the e-learning tutorials were created. Each initial draft was revised several times based on interviews with introductory physics students and feedback from graduate students and several professors who were asked to work through them and provide feedback to ensure that they were comfortable with the wording of the sub-problems and progression of the tutorial. During this refinement process, the fine-tuned versions of the e-learning tutorials were implemented in one-on-one think aloud [24, 25] interviews with introductory physics students and were shown to improve student performance on the paired problems that were developed in parallel with the tutorials.

Comparing the lengths of the three e-learning tutorials, we note that the Newton's second law and conservation of energy/work-energy theorem tutorials were made up of 17 and 19 sub-problems, respectively, while the conservation of angular momentum tutorial is made up of 7 sub-problems. Based upon prior research, it is possible that the more complex problems may be more effective in helping students learn a systematic approach to problem solving and connect different physics concepts [26]. Newton's second law was complex because it involved several blocks and conservation of energy/work-energy problem was complex because it was a context-rich problem and involved two physics principles. However, since angular momentum conservation is challenging to conceptualize, the researchers collectively decided to investigate how a short tutorial that focuses on why angular momentum is conserved in a given context and how to apply the angular momentum conservation helps students transfer their learning to an isomorphic paired problem.

## 3. RESEARCH METHODOLOGY

Below, we describe the methodology for the implementation of the three e-learning tutorials in one-on-one implementation with student volunteers and as a self-study tool as part of traditional brick and mortar large introductory physics courses at the University of Pittsburgh (which is a large, typical state-affiliated university in the US) to evaluate their effectiveness.

### 3.1. Deliberate One-on-one implementation

One of our goals was to determine the effectiveness of interactive e-learning tutorials in one-on-one implementation. Therefore, they were administered individually to introductory students in deliberate one-on-one think aloud interview settings [27] so that researchers could monitor whether students were using them as intended. These students were paid volunteers who responded to a flyer distributed in the introductory physics classes and already had traditional classroom instruction related to physics concepts covered in the tutorial. These interviews were audio-recorded.

In this deliberate one-on-one implementation, students were observed by a researcher as they worked on the e-learning tutorials but they otherwise followed the same instructions that were given to the students in the large introductory physics courses, who used them as a self-study tool. However, in one-on-one implementation, the researchers made sure that students adhered to the guidelines. For example, students had to first attempt to outline the solution to the tutorial problem to the best of their ability and only then asked to start the tutorial and attempted each sub-problem in the appropriate order. As noted, throughout this one-on-one implementation process in which a student worked on the e-learning tutorials, the student was asked to think aloud while being audio-recorded and a researcher made further record of his observations of each student's interaction with the e-learning tool. This process was repeated with each student for each tutorial.

Twenty-two 2-3 hour long, one-on-one, think-aloud interviews were conducted with volunteers who were either in an algebra or calculus-based introductory physics course. We note that although several interviews were conducted with students individually during the development of the e-learning tutorials, we conducted these 22 additional one-on-one interviews with students who had been exposed to a typical classroom treatment of Newton's second law, conservation of energy/work energy theorem, and conservation of angular momentum. In 17 of these interviews three tutorials were covered and in the remaining 5 only two tutorials were covered due to the interviewed student working slowly. In each case, the order in which the tutorials were presented was changed resulting in approximately 20 individual interviews for each tutorial (half of which were with students in the algebra-based physics course and the other half were with students in the calculus-based physics course). Throughout this process, a researcher was present to provide materials, and the interviewer ensured that the students explicitly followed the instructions provided and made effective use of the tutorial. Students were asked to think aloud so that researcher could understand their thought processes and the researcher remained silent while the students worked unless they became quiet, in which case the researcher prompted students to keep talking. After working through the entire e-learning tutorial, the students worked on the corresponding paired problem.

3.2 Large scale implementation of the tutorial as a self-study tool

We also investigated the effectiveness of the e-learning tutorials for a diverse group of introductory physics students in brick-and-mortar introductory physics courses in which researchers provided guidelines but otherwise had no control over how the tutorials were used by the students. The tutorials were implemented as self-study tools in two traditional, large introductory physics courses. The first course was an algebra-based first semester introductory physics course with roughly 385 students (split into two sections). These students came from varied backgrounds in math and science with a majority of them pursuing bioscience or neuroscience majors. The second course was a calculus-based first semester introductory physics course with roughly 350 students (also split into two sections). The students in this course were almost entirely physical science, mathematics and engineering majors.

Each of the three tutorials was posted on the course website as a self-study tool after students had received classroom instruction in relevant concepts. They were intended to be used at students' discretion after the associated physics concepts and principles were introduced in lecture but before students had the opportunity to do the associated homework problems. The links to the tutorial were uploaded on the course website but the amount of time each student spent working through them could not be tracked. Students were aware that no points would be awarded for completing the e-learning tutorials, but announcements were made in class, posted on the course website, and sent via email informing students that the tutorials were available when relevant concepts were covered in class. The incentive that the instructors gave to their students for engaging with these self-paced tutorials was that they would be helpful for solving assigned homework problems and in-class quiz problems (paired problems) for that week.

The paired problems associated with each tutorial were given to students during their regular weekly recitation class. These quizzes with paired problems were given after students had been given access to the associated tutorial for an entire week. Each paired problem was administered in the week following instruction in a particular concept. All students had sufficient time to complete the quizzes. Students were given a grade based on their performance on these paired problems as their weekly quiz grade. At the top of each of the paired problem quizzes administered in the recitation, students were asked the following questions and assured that the answers to these questions would not influence their grade:
- Have you worked on the corresponding online tutorial?
- Was the tutorial effective at clarifying any issues you had with the problem covered in the tutorial?
- If the tutorial was ineffective, explain what can be done to make it effective?

- How much time did you spend on the tutorial?

To compare the performance of the students who worked on the e-learning tutorials in a one-on-one interview vs. those who used them as a self-study tool, we compared student performance on the paired problems. Rubrics were developed by three graduate students and a professor for each of the paired problems. Once the rubric for grading each paired problem was agreed upon, 10% of the paired problem quizzes were graded independently by three graduate students and a professor with the finalized version of the rubric. When the scores were compared, the inter-rater agreement was better than 90% across all graders.

## 4. RESULTS

### 4.1 Deliberate one-on-one implementation

The purpose of conducting the 22 individual interviews was to gauge the effectiveness of these e-learning tutorials when administered in a controlled environment in which a researcher can monitor that they are being used as intended compared to their use as a self-study tool in large enrollment classes. Table 1 shows the average performance of students along with the standard deviation in the one-on-one interview group. Table 1 shows that students in a one-on-one interview setting had an average score of above 80% on all the paired problems.

**Table 1.** The average paired problem scores and standard deviations (SD) for students in the one-on-one implementation group.

| Physics Principle | Interview Group (SD) | Number of Students (Algebra, Calculus) |
|---|---|---|
| Newton's Second Law | 86.0% (15.9%) | 20 (11, 9) |
| Conservation of Energy/Work-Energy Theorem | 95.5% (11.8%) | 21 (11, 10) |
| Conservation of Angular Momentum | 83.3% (16.0%) | 20 (12, 8) |

### 4.2 Large scale implementation of the e-learning tutorial as a self-study tool

In this section, we discuss the average performance of students on the paired problems in the large scale implementation of the tutorials as a self-study tool. Before working on the paired problem, students were asked whether they had worked on the tutorial as a self-study tool and how much time they spent working on it. Also, students were asked to write down on the paired problem if the tutorial was effective at clarifying any issues they had with the tutorial problem. They were told that they should be honest because the answer to the question would not impact their grade. Only approximately 60% or less of the students reported that they worked through each of the tutorials in both the calculus-based and algebra-based classes. Table 2 shows that a majority of students thought that the tutorials were effective at clarifying issues they had with the problem. Students were also asked to write down whether anything can be done to make the tutorials effective if they thought it was ineffective. Most students provided no comments and a few students who provided comments generally noted that perhaps they can be made shorter so that they can quickly browse over them.

**Table 2**. Student responses to the question "Was the tutorial effective at clarifying any issues you had with the problem covered in the tutorial?" in large enrollment classes as a self-study tool

|  | Yes | No | No Response |
|---|---|---|---|
| Newton's Second (Algebra) | 76 | 7 | 4 |
| Newton's Second (Calculus) | 135 | 11 | 5 |
| Conservation of Energy (Algebra) | 168 | 17 | 13 |
| Conservation of Energy (Calculus) | 139 | 19 | 7 |
| Conservation of Angular Momentum (Algebra) | 169 | 16 | 2 |
| Conservation of Angular Momentum (Calculus) | 121 | 22 | 7 |

In Table 3, students in the large introductory physics classes are divided into the "tutorial" or "non-tutorial" group based upon self-reported data about whether they worked on the tutorial regardless of how much time they had spent working on it. Table 3 shows that two of the three tutorials that were given as a self-study tool resulted in a statistically significant improvement in student performance (compared to the non-tutorial group) on the paired problem in the algebra-based group. Only one of the tutorials resulted in a statistically significant improvement in the calculus-based group. The only e-learning tutorial that resulted in a statistically significant improvement for both the algebra-based and calculus-based groups is the conservation of angular momentum tutorial. Table 3 also shows the Hake gain [28], which is (posttest % − pretest %)/(100% − pretest %). This Hake gain between the tutorial and non-tutorial groups is low for all tutorials for both algebra-based and calculus-based classes except for the conversation of angular momentum tutorial in a calculus-based class (which is 0.369).

**Table 3.** Comparison of the performance on the paired problem and standard deviations (SD) for students who used the tutorials and those who did not use them in large, brick-and-mortar introductory physics courses.

**Algebra-Based Group**

| Tutorial | Tutorial Group Average (SD) | Number of students | Non-Tutorial Group Average (SD) | Number of students | p-value | Hake Gain |
|---|---|---|---|---|---|---|
| Newton's Second Law | 53.9% (29.2%) | 87 | 44.6% (29.1%) | 274 | 0.001 | 0.167 |
| Conservation of Energy | 46.9% (35.2%) | 165 | 41.5% (38.2%) | 172 | 0.178 | 0.092 |
| Conservation of Angular Momentum | 53.9% (29.4%) | 150 | 44.0% (31.9%) | 186 | 0.003 | 0.177 |

**Calculus-Based Group**

| Tutorial | Tutorial Group Average (SD) | Number of students | Non-Tutorial Group Average (SD) | Number of students | p-value | Hake Gain |
|---|---|---|---|---|---|---|
| Newton's Second Law | 77.5% (27.1%) | 135 | 72.8% (28.9%) | 197 | 0.142 | 0.173 |
| Conservation of Energy | 81.8% (27.5%) | 185 | 78.8% (32.2%) | 133 | 0.385 | 0.142 |
| Conservation of Angular Momentum | 69.1% (26.2%) | 184 | 51.0% (29.9%) | 115 | <0.001 | 0.369 |

Table 4 compares the performance of students in one-on-one implementation of the e-learning tutorials with those who claimed they had used them as a self-study tool. Table 4 shows a considerably higher average score for students in the one-on-one implementation group compared to those in the large scale implementation as a self-study group for all three tutorials. A noteworthy observation is that the one-on-one implementation group, composed of 12 students from large enrollment algebra-based courses and 10 students from calculus based courses, scored considerably higher than both large scale self-study implementation groups. We do not separate the algebra-based and calculus-based groups in Table 4 since there were only 22 students including both groups in the one-on-one implementation group.

**Table 4.** Comparison of the average paired problem scores and standard deviations (SD) for students in the one-on-one interview group as compared to those who made use of the tutorial in the large enrollment classes in the self-study group.

| Physics Principle | Interview Group (SD) | Calculus-Based Self-Study Implementation (SD) | Algebra-Based Self-Study Implementation (SD) |
|---|---|---|---|
| Newton's Second Law | 86.0% (15.9%) | 77.5% (27.1%) | 53.9% (29.2%) |
| Conservation of Energy | 95.5% (11.8%) | 81.8% (27.5%) | 46.9% (35.2%) |
| Conservation of Angular Momentum | 83.3% (16.0%) | 69.1% (26.2%) | 53.9% (29.4%) |

## 5. DISCUSSION

We evaluated the relative effectiveness of the research-based e-learning tutorials when students worked on them as a self-study tool at their own discretion without supervision in large enrollment introductory physics classes as compared to in a deliberate one-on-one setting. Students making use of the tutorials as a deliberate one-on-one tool worked on them with a researcher monitoring the students so that they used them as prescribed. The students were prompted to think aloud while working on them but otherwise were not disturbed. In the self-study implementation, although students were instructed to follow the same guidelines for effective learning, they could potentially take a short cut and skip sub-problems if they decided not to adopt a deliberate learning approach while using these research-based tools. We found that many students in the self-study implementation group did not work through the tutorials, and if they did, the improvement in their average performance is not impressive with respect to the gains expected from these research-based e-learning tutorials (evidenced by the performance on the paired problems of those in the individual one-on-one implementation group). This dichotomy between the performance of the self-study group and the one-on-one implementation group suggests that research-based tutorials, when used as intended, can be an excellent learning tool for introductory physics students across diverse levels of prior preparation, experience and mathematical background but getting students to engage with them effectively as a self-study tool can be challenging.

One possible reason for the significantly better performance on paired problems among the one-on-one implementation group as compared to either the algebra-based or calculus-based self-study groups is the ineffective approaches to using the tutorial as a self-study tool. As noted, students were made aware (by way of e-mails, announcements on the course web-page and a description of the tutorials given to them verbally during their regularly scheduled class time) that working on the self-paced interactive tutorials posted on the course website does not contribute directly to the grade but that working through them deliberately will help them learn and improve their homework and quiz performance. Upon examining student comments and other data gathered with their response to the paired problems in the self-study group, it appears that some students who claimed to make use of the tutorials may not have used them effectively. Some students explicitly commented that they "skimmed" or "looked over" the tutorials but that type of engagement with the e-learning tool may not help them learn. A detailed look at the performance of students enrolled in the introductory physics courses on the paired problems indeed suggests that many students may have memorized certain equations by browsing over the tutorials, expecting that those equations may help them in solving the in-class quiz problems, instead of engaging with the self-paced tools as instructed in a systematic manner. Interestingly, in a survey given at the end of the course to students who used them as a self-study tool, a majority noted that they thought that the tutorials were effective even though their performance on the paired problems reflected that they had not learned significantly from them.

Another reason for the significantly better performance on paired problems among the one-on-one implementation group as compared to either the algebra-based or calculus-based self-study groups is the potentially inaccurate self-reporting of how much time students spent working on them. Many students' self-reported time spent working on the tutorials appears to be inconsistent with the time taken by students who worked on them in one-on-one implementation. For many students in the self-study group, the self-reported time spent working on the tutorial is often considerably lower than the time taken by the quickest students in deliberate one-on-one situation. On the other hand, some students in the self-study group reported that working on the tutorial took them a significantly longer time than the typical time taken by the 22 students in the deliberate one-on-one implementation group. For example, the conservation of angular momentum e-learning tutorial took most interviewed students 15-30 minutes but some students in the self-study group reported spending up to 1.5 hours on this tutorial. The conservation of angular momentum tutorial was the shortest tutorial (made up of 7 sub-problems), while the Newton's second law and conservation of energy tutorials were made up of 17 and 19 sub-problems, respectively. The conservation of angular momentum tutorial was the only tutorial that resulted in a statistically significant improvement for both the algebra-based and calculus-based students when used as a self-study tool in large introductory physics courses. This shorter tutorial on the conservation of angular momentum only involved application of one physics principle in one situation so students may have found it easier to follow. However, the students in the one-on-one interview setting had an average score of over 80% on the angular momentum paired problem. This score is 45% higher than the average score of students in the self-study group in brick-and-mortar introductory physics courses. This dichotomy indicates that students in the self-study group may have engaged with the tutorials in a manner not conducive to learning. These students may be lacking self-regulation, time-management skills, focus, etc. while working through the tutorial.

It appears that without sufficient support to help students develop self-management and time-management skills and incentives to motivate students to engage with the self-paced tutorials, many students may not follow the guidelines for effectively using them. The haphazard use of these research-based self-paced tools can reduce their effectiveness significantly. The significantly lower performance of students in the self-study group in this investigation supports the notion that major challenges in implementing research-based tutorials as self-study tools are likely to be issues such as students' level of motivation, self-regulation, and time management to engage with them [29-31]. Many students have difficulty internalizing that much of the value to be gained from these self-paced tools depends on them interacting with them in a prescribed manner. For example, students who explicitly reported having "skimmed" through the tutorials most likely did not engage with each of the individual sub-problems as they were prescribed to do. Additionally, they may not have attempted to solve the tutorial problem on their own without the scaffolding provided by the tutorial as they were asked to do before starting to work on the tutorial. Although the instructions for effective usage of these self-study tools were provided to students through several channels, many students may have interacted with the self-study tools only superficially. Even among students observed in deliberate one-on-one interviews, some had to be prompted several times to make a prediction for each sub-problem and articulate their reason for selecting an answer before selecting an answer rather than randomly guessing an answer.

The fact that the tutorials were ineffective as a self-study tool (even though they were effective in deliberate, one-on-one administration for a variety of students) attests to the difficulty in making any research-based online learning tool effective for a diverse group of students. In fact, since the sharp increase in the availability of e-learning learning tools in the last decade, the initial mantra of policy makers, media and public at large was that soon everybody will have the opportunity to learn any subject via inexpensive, self-paced, e-learning tools without attending brick-and-mortar classes [32, 33]. The development of e-learning tools was hailed as a panacea for educating a diverse group of students with different backgrounds, including those with very different prior preparations and skills, and lacking significant financial resources. However, data from the Massive Open Online Courses (MOOCs) suggest that while a variety of people enroll in MOOCs, a majority of those who complete them typically have a predictable profile [34-36]. In particular, a majority of those who complete the MOOCs already have a college degree and many even have a graduate degree. Research on MOOCs suggests that the students from low socio-economic backgrounds, who do not have the money to pay for traditional college (for whom the MOOCs were originally intended as an altruistic act by the institutions of higher education), are very likely to drop out within the first few weeks. This attrition may partly be due to students lacking sufficient motivation, discipline, self-regulation, and time-management skills to engage effectively with the e-learning tools. This hypothesis is supported by the fact that a majority of those who complete the MOOCs already possess college degrees, and are likely to have superior motivation and skills to take advantage of the self-study tools compared to those who drop out of the MOOCs.

In fact, not only do MOOCs and online courses use e-learning tools, but hybrid or blended-courses also use them to varying degrees. Moreover, the instructors in a typical brick and mortar course often integrate web-based components as learning tools similar to the study described here. Many instructors realize that self-paced, out of class learning tools are critical even for a traditional, brick-and-mortar course, especially since the self-paced tools address some of the challenges involved in educating students with diverse motivation, prior preparation and backgrounds in a given course [37-41]. These instructors may aim the lectures at an average student in the course and assume that the students below the class average at a given time will catch up using the self-study tools instructors prescribe. However, many students may not engage with them as prescribed, as suggested by our investigation.

Apart from the self-regulation and time-management skills necessary to hold oneself accountable for learning from the self-study tools, motivation can play a critical role in whether students take advantage of these tools. One may hypothesize that since MOOCS are free, students may have less motivation to learn from them compared to the self-study tools prescribed in a course they are paying for. However, the study discussed here suggests that many students who are paying for their classes may not have the self-regulation and time-management skills and other skills to engage with the self-study tools unless they are provided sufficient extrinsic motivation and support to learn via those tools. It is therefore important to contemplate different facets of learning from self-paced learning tools in order to provide appropriate support and incentives to students to benefit from them, especially if they are research-based tools that have been found effective in one-on-one implementation. Below, we propose a theoretical framework for this purpose.

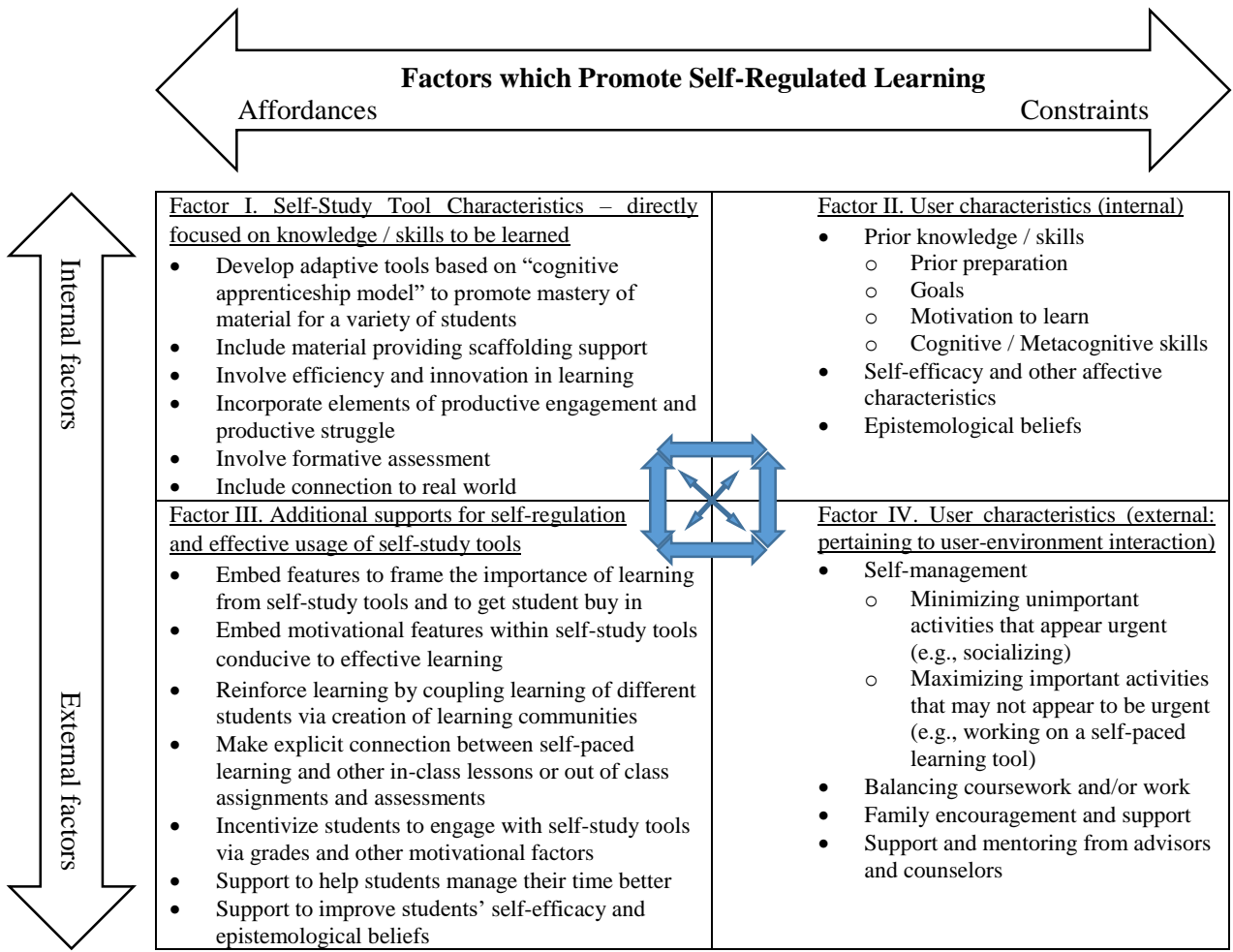

Figure 3: Self-study for Engaged Learning Framework (SELF)

The framework, which we call the Self-study for Engaged Learning Framework (SELF), is proposed to help students with diverse backgrounds benefit from self-study tools. Our framework consists of four quadrants as shown in Figure 3 and all of them must be considered holistically in order to help a diverse group of students learn effectively from self-study tools. The left two quadrants focus on the internal and external affordances for learning via self-study tools and the right two quadrants focus on the constraints imposed on the effective usage of self-study tools by user characteristics (internal) and the characteristics of the user-environment interaction (external). The constraints in Figure 3 in the right two quadrants refer to whether the user characteristics and the characteristics of user-environment interaction are conducive to helping students learn from the self-study tools. Affordances [42-46] are defined as qualities of systems, which can be harnessed appropriately to support effective interactions between individuals and tools in different situations. Here, we use the word "affordance" to imply both the internal and external features of a self-study tool that afford learning for students with diverse backgrounds and characteristics (see the left two quadrants in Figure 3). This type of classification of affordances into internal and external categories is common in other contexts as well since making good use of both the internal and external affordances is critical for success. For example, in order to land a plane safely, the internal affordances could refer to the features of the plane itself while the external affordances could refer to the features of the runway, and weather on a particular day when a pilot is landing the plane.

Most of the research-based self-study e-learning tools developed so far have mainly focused on the upper two quadrants of the framework in Fig. 3. In the upper-left quadrant, the focus is on the self-study tool characteristics that directly focus on knowledge and skills to be learned via the self-study tool. The cognitive apprenticeship model [20] can be used to develop adaptive self-study tools to promote mastery of the material for a variety of students [47]. These materials, when developed carefully via research in education, can provide scaffolding support to a variety of students. In order to make the self-study tools effective, educators often consider the user characteristics in the upper-right quadrant [48-55]. The various models of learning lead to similar conclusions about how to connect user characteristics with the characteristics of the self-study tools (i.e., how to connect Factors I and II). For example, Schwartz, Bransford, and Sears' [56] preparation for future learning model emphasizes that in order for students to engage appropriately with learning tools, there should be elements of both efficiency and innovation embedded in the instructional tools and design. One interpretation of this model is that if the students are asked to engage with learning tools that are too efficient, they will get bored and disengage. On the other hand, if the learning tools are too innovative, students will struggle so much while engaging with them that they will get frustrated and give up. Thus, the learning tools should be adaptive in that they blend elements of both efficiency and innovation, allowing students to engage and struggle productively while learning [57, 58]. In addition, effective self-study tools should have formative assessment built into them so that students can receive feedback and evaluate their own learning as they make progress. Since student characteristics within a particular class vary, carefully designed adaptive interactive self-study tools can provide appropriate balance of innovation and efficiency for a variety of students [59-62]. Students who are lacking some elements of prior knowledge can benefit from a carefully designed self-study tool which involves formative assessment, allows students to make mistakes but learn from them and try again, and scaffolds their learning by providing elements of both efficiency and innovation [63-70].

In the study described here, the research-based tutorials included considerations of Factors I and II. For example, the tutorials provided an opportunity for productive struggle - they specifically encouraged students to work on each tutorial problem before starting to work on each of the sub-problems. The act of struggling with the tutorial problem can help students connect what they are learning with their prior knowledge and aid in learning. Additionally, struggling with the tutorial problem before engaging with the tutorial may increase students' motivation to engage deliberately with the e-learning tool as prescribed. However, the e-learning tutorials could also be improved based upon consideration of Factors I and II. For example, the longer e-learning tutorials were more complex since they either involved application of more than one physics principle or application of the same principle (Newton's second law) in different contexts. These longer e-learning tutorials are useful for helping students develop both content knowledge and skills to solve complex problems. However, since many students may have disengaged with the longer tutorials while using them as a self-study tool, finding better ways to keep students motivated throughout while working through them should be a high priority rather than only developing shorter tutorials focused on one physics concept/principle [71]. Based upon considerations of Factors I and II, one strategy that may make them more effective is to break the multi-principle tutorials into single-principle tutorials. After working through the single-principle tutorials, students can then work through a subsequent multi-principle tutorial that combines the learning in those single-concept tutorials. Since students would have been exposed to the individual concepts in various single-concept tutorials, they will be more likely to effectively engage with the multi-principle tutorial that consolidates those principles into a more complex problem.

However, research-based self-paced learning tools which take into account students' prior knowledge will not necessarily help them learn if students do not take advantage of the self-study tools to learn in an effective manner. While the top two quadrants in our framework are often considered in the development of self-study tools, the lower two quadrants of our theoretical framework for learning using self-study tools have mostly been ignored while designing and implementing most self-study tools even if those tools are carefully developed. But as the study presented here points out, these lower two quadrants are likely to play a critical role in whether students, who are especially in need of remediation via self-study tools, take advantage of these self-study tools. The lower right quadrant or Factor IV focuses on external student-environment interaction characteristics, e.g., how students interact with their surroundings and how they manage their time and regulate themselves. For example, if the students get bogged down with unimportant activities (e.g., communicating with friends on social media), they are unlikely to make time for activities that are important (e.g., learning from self-study tools). Factor IV also involves support that students may receive from their environments such as help from family, advisors, mentors and counsellors to manage their time better and engage in learning using self-study tools effectively. In our study, students' engagement with the self-study tools may have been impacted by whether they have self-management skills, time-management skills, family encouragement, and support from advisors and counselors.

The question then boils down to whether there are external affordances that can be provided during the implementation of the self-study tools to assist students who otherwise may not engage with them effectively due to personal constraints. This external additional support from educators for self-regulation and effective use of the self-study tool is included in the lower left quadrant (Factor III) and focuses on providing motivation and support for engagement, taking into account the user characteristics and user-environment interactions. Consideration of the various types of support in quadrant III during the implementation of the self-study tools is critical to ensure that most students engage with the self-study tools effectively. In our study, students may have engaged more effectively with the e-learning tutorials if elements from Factor III were included in the implementation of the self-study tools. For example, self-study tool developers or implementers can consider embedding modules that focus on motivating students to engage with the self-study tools effectively and strive to get buy-in from students by having them think carefully about why they should engage effectively with these tools and how they can help them in the long term.

Similarly, the students who are struggling to manage their time well can be provided some modules to guide them in making a better daily schedule which includes time to learn from the self-study tools (once students have made a schedule that includes time slots for learning from self-study tools, electronic notifications can remind them of their schedule as needed). In addition, making explicit connection between self-paced learning and other in-class lessons or out of class assignments and assessments can also help students engage with the self-study tools more effectively.

Moreover, students who have difficulty engaging with the self-study tools due to lack of self-efficacy or unproductive epistemological beliefs [72] about learning can be guided to help them develop self-efficacy [73, 74] and productive epistemological beliefs. For example, a short online intervention has been shown to improve student self-efficacy significantly [75]. Similarly, students who have unproductive epistemological beliefs (such as physics is just a collection of facts and formulas, only a few smart people can do physics, and they should just memorize physics formulas and regurgitate them) are unlikely to productively engage with the self-study tools designed to help them develop expertise in physics. It is important to address these issues in order to ensure that students who are most in need of learning using self-study tools actually benefit from them and retain what they learn [76-86].

Another factor (see Factor III of the framework) that may help students engage with these tools effectively is participation in learning communities of students who are all expected to learn from the self-study tools and then have them engage in some follow up activities in a group environment (this group work can be done online or in-person depending on the constraints of the class). In this way, individual students may feel more accountable to their group members and effectively use self-study activities to prepare for the group activities. For example, in the study discussed here, encouraging and incentivizing students to work in these types of learning communities could have aided students in engaging with the self-paced e-learning tutorials more effectively. In particular, if students knew that they were assigned to work with a group on a complex physics problem, they may have had more motivation to work through the e-learning tutorials individually in order to prepare for the group work.

Moreover, having more effective grade incentives [87, 88] to learn from the self-study tools is another external factor that can also increase student engagement (see Factor III of the framework). For example, to help students engage effectively with the e-learning tutorials, an instructor could incentivize participation in learning via grade incentives to ensure that students work on them as prescribed. Also, if students work systematically on them and are engaged throughout, they are unlikely to have cognitive overload [89, 90] since learning is scaffolded throughout and one sub-problem builds on another. One motivating factor would be to award course credit to students based on their answers to each sub-problem with decreasing score if they guess multiple times. This strategy might be more successful at motivating them to answer each sub-problem carefully (as opposed to randomly guessing an answer) while working through the e-learning tutorial. In addition, it is possible that if students in the study described here were asked to submit a copy of their answers to each sub-problem of the e-learning tutorial and explain why each alternative choice to each sub-problem is incorrect as part of their homework, it may have increased their motivation to engage with these self-study tools (especially because students have many conflicting priorities for their time and they may not engage with self-study tools if working through them is not *directly* tied to the grade).

We note that in our framework, Factor III may also impact Factor IV. When students are motivated to think about the importance of using self-study tools, given credit to work through the self-study tools, work in learning communities that keep each student accountable while providing mutual support, and can discern the connections between the self-study tools and in-class assignments, homework, and quizzes, they may manage their time more effectively. Connecting self-study tool content to real-world applications can also increase student motivation to learn from these self-paced tools. It is also important to note that Factors I and III can impact Factors II and IV so we cannot disentangle any of these factors. Students who are lacking prior preparation may also have difficulty in managing their time effectively. But there are often students who are prepared to learn using self-study tools but lack time management skills. Other students may not have good prior preparation but they may have good time management skills. In all these cases, in order to help students learn effectively from the self-study tools, Factor I and Factor III (the affordances) should outweigh the constraints. Therefore, consideration of Factor III, which is often ignored by educators developing self-study tools, is critical.

## 6. SUMMARY AND CONCLUSIONS

We compare the effectiveness of three e-learning tutorials when used as a self-study tool in large enrollment classes with their effectiveness when they were used by students in a closely monitored, deliberate, one-on-one setting. At the beginning of these tutorials, students are asked to attempt to solve the problem themselves. Then, if necessary, they solve each sub-problem and select the appropriate choice for the multiple choice question that they feel best fits the answer to a sub-problem. The tutorials are adaptive in that if students' answers to the multiple-choice questions are incorrect they are directed to obtain suitable help commensurate with their difficulty. If students answer a multiple-choice sub-problem incorrectly, they are returned to the same sub-problem to make another attempt at solving it after receiving help. If they answer a sub-problem correctly, they encounter a reinforcing statement about why that option is correct before proceeding to the next multiple-choice sub-problem.

The students who used the e-learning tutorials in a one-on-one setting were instructed to work on them following the same instructions as those provided to students using these tutorials as a self-study tool, but they were monitored, i.e., they had to follow the instructions and could not skip any part. After working on the tutorial (either as a self-study tool or as a deliberate one-on-one e-learning tool), students' knowledge of the associated physics concepts was evaluated via their performance on an associated paired problem that involves the same physics principle/concepts as the e-learning tutorial problem. We find that students in the deliberate, one-on-one implementation group significantly outperformed those in the self-study group on the paired problems.

The fact that students had to follow the correct protocol in the deliberate one-on-one implementation (i.e., start by solving the problem without any help and then work on the tutorial as intended) may have contributed to the success of the tutorials in deliberate one-on-one interviews. On the other hand, the lack of effectiveness when students used the tutorials as a self-study tool is likely due to students engaging with the tutorial in ways other than those outlined for them. Despite the encouragement from the instructor, it was difficult to ensure that students used an effective approach while working through the self-paced e-learning tutorials at their own time. If students bypass the step of first attempting the tutorial problem and do not engage with the tutorial deliberately (i.e., thinking about and attempting to answer each sub-problem before looking at the scaffolding and feedback), the effectiveness of the tutorial can be greatly diminished. Each step in an e-learning tutorial is designed to help students recognize and resolve any difficulties that they have while strengthening their knowledge of physics concepts as they work on them. Students who did not use this deliberate approach outlined for them when engaging with self-study tools are unlikely to benefit much from them.

Our investigation suggests that despite the ease with which students can access these adaptive e-learning tutorials, there are challenges in ensuring that students, especially those who need out-of-class scaffolding support [91, 92], use them effectively as a self-study tool as intended. The implications of these findings may extend to other self-study tools. In particular, students interacting with even the best designed self-study tools are likely to do so in ways other than those prescribed explicitly, which can greatly diminish the tool's effectiveness. This limitation is inherent to self-study tools that have no means of regulating the ways in which the student interacts with them unless issues discussed in our framework in the lower left quadrant are explicitly incorporated.

A lack of sufficient motivation, discipline, self-regulation, and time-management skills while engaging in learning using self-study tools may turn out to be the biggest impediment in implementing research-based e-learning tools for use as self-study tools. The theoretical framework we propose emphasizes that in order for students with diverse backgrounds and prior preparations to benefit from self-study tools, educators must holistically consider various facets of student engagement with self-study tools and incorporate them in their development and implementation of those tools.

**Acknowledgements**

We thank the US National Science Foundation for financial support and members of the University of Pittsburgh physics education research group for their help with the rubric development and grading to establish inter-rater reliability.